\documentclass[preprintnumbers,superscriptaddress,nofootinbib,aps,prd,floatfix]{revtex4}

\usepackage{graphicx}  
\usepackage{dcolumn}   
\usepackage{bm}        
\usepackage{amssymb,amsmath}
\usepackage{wasysym}
\usepackage{adjustbox,lipsum}
\usepackage{subfigure}
\usepackage{hyperref}

\usepackage[usenames,dvipsnames,svgnames]{xcolor}
\usepackage{hyperref}
\hypersetup{colorlinks, citecolor=black, linkcolor=black, urlcolor=black}

\usepackage{textcomp}
\usepackage{amsmath}
\usepackage{amsfonts}
\usepackage{amssymb}
\usepackage{bbm}                
\usepackage{graphicx, rotating}
\usepackage{epstopdf}
\usepackage{epsfig}
\usepackage{latexsym}
\usepackage{graphicx}
\usepackage{bbold}
\usepackage{color}
\usepackage{amsmath,amssymb}
\usepackage{slashed}

\newcommand{\beq}{\begin{eqnarray}}
\newcommand{\eeq}{\end{eqnarray}}

\def\ltap{\ \raise.3ex\hbox{$<$\kern-.75em\lower1ex\hbox{$\sim$}}\ }
\def\gtap{\ \raise.3ex\hbox{$>$\kern-.75em\lower1ex\hbox{$\sim$}}\ }

\def\be{\begin{equation}}
\def\ee{\end{equation}}
\def\bea{\begin{eqnarray}}
\def\eea{\end{eqnarray}}

\newcommand{\hc}{\rm h.c.}
\definecolor{red1}{cmyk}{0,1,1,0.3}

\preprint{IPPP/19/29}

\begin{document}
\title{Revisiting the $t\bar{t}hh$ channel at the FCC-hh}
\author{Shankha Banerjee}\email{shankha.banerjee@durham.ac.uk}
\affiliation{Institute of Particle Physics Phenomenology, Durham University, Durham DH1 3LE, UK}
\author{Frank Krauss}\email{frank.krauss@durham.ac.uk}
\affiliation{Institute of Particle Physics Phenomenology, Durham University, Durham DH1 3LE, UK}
\author{Michael Spannowsky}\email{michael.spannowsky@durham.ac.uk}
\affiliation{Institute of Particle Physics Phenomenology, Durham University, Durham DH1 3LE, UK}


\begin{abstract}
The exploration of the scalar sector of the Standard Model is at the core of current and future science programs at collider experiments, with increasing focus on the self-interaction of the Higgs boson.  This important parameter of the Higgs sector can be measured in various channels, among the production of a Higgs boson associated with a top-quark pair, $\bar{t}thh$.  In this paper we study this channel and its potential to measure or constrain the self-coupling and possible new physics contributions at a future 100 TeV proton-proton collider. Analysing this highly complex final state adds to the sensitivity for enhanced self-coupling interactions, and we argue that a measurement of this process is a necessity to constrain blind directions in the multi-dimensional parameter space of well-motivated new physics scenarios.
\end{abstract}
\maketitle

\section{Introduction}
\label{sec:1}

After the discovery of the Higgs boson~\cite{Aad:2012tfa, Chatrchyan:2012xdj} with a mass of $m_h \simeq 125$ GeV, the focus of contemporary and future particle phenomenology shifted to the determination of its properties. Early on, spin and CP properties have already been found to be in extremely good agreement with Standard Model expectations~\cite{Aad:2015mxa, Aad:2016nal, Sirunyan:2017tqd, Khachatryan:2016tnr}, and the couplings of the Higgs boson to other particles, in particular the heavy gauge bosons and the third-generation fermions, are increasingly precisely measured~\cite{Khachatryan:2016vau, Sirunyan:2018kst, Aaboud:2018urx}, thereby reducing the parameter space for extensions of the Standard Model~\cite{Englert:2015hrx,deBlas:2016ojx,Butter:2016cvz,Ellis:2018gqa}.  This leaves the form and parameters of the Higgs potential, and in particular the Higgs self-coupling, as the experimentally least constrained sector of the Standard Model. It is therefore not surprising that measurements of or constraints to the triple-Higgs coupling are one of the center pieces of ongoing efforts for the high-luminosity run of the LHC and an important part of particle phenomenology at possible future collider experiments. If the Higgs self-coupling is the only modification to the Standard Model, various ways have been proposed at existing and future colliders to search for this interaction. These approaches can be classified into three categories: in processes without Higgs bosons in the final state, electroweak precision observables can set a limit to $\lambda_{hhh}$~\cite{vanderBij:1985ww, Degrassi:2017ucl, Kribs:2017znd}; higher-order corrections in single-Higgs production processes~\cite{Bizon:2016wgr, Degrassi:2016wml, Maltoni:2017ims} constrain the Higgs self-coupling; and double-Higgs production processes will provide direct sensitivity on this coupling in upcoming LHC and possible future high-energy collider runs~\cite{Baur:2002rb, Baur:2002qd, Dolan:2012rv, Baglio:2012np, Cao:2016zob,  Adhikary:2017jtu, Benedikt:2018csr, CMS:2018ccd}, while the latter are expected to provide the best sensitivity on $\lambda_{hhh}$ during the LHC's high-luminosity runs.

Within the class of multi-Higgs production processes, the overwhelming focus to date was directed towards the channel with the largest cross section, \textit{i.e.} Higgs-boson pair production in gluon fusion, while other channels, such as Higgs-pair production in association with other particles, \textit{e.g.}\ $pp\to hhjj$~\cite{Dolan:2013rja, Dolan:2015zja,Bishara:2016kjn} or $pp\to t\bar{t}hh$~\cite{Englert:2014uqa,Liu:2014rva}, have been somewhat neglected. In Refs.~\cite{Englert:2014uqa,Liu:2014rva} it has been found that the $t\bar{t}hh$ channel at the $14$ TeV high-luminosity run of the LHC may provide welcome additional statistical power for a determination of the trilinear Higgs coupling. The feature that sets this channel apart from the gluon-induced Higgs-pair production process or the weak-boson induced production of $hhjj$ arises due to the absence of a reduced cross section for large values of $\lambda_{hhh}$~\cite{Frederix:2014hta}. Thus, $t\bar{t}hh$ could be particularly useful in setting a stringent limit to enhanced self-interactions of the Higgs boson.

A further motivation to measure $t\bar{t}hh$ final states arises when modifications of Higgs interactions originate in models where the Higgs field is realised in a non-linear way, \textit{e.g.}\ composite Higgs models~\cite{Kaplan:1983sm,Panico:2015jxa,Chala:2018ari}. There, the $t\bar{t}h$ and $t\bar{t}hh$ couplings are decorrelated~\cite{Grober:2010yv,Gillioz:2012se}, leading to a blind direction in the parameter space of effective operators when only probing them through the top-associated single Higgs production process, $pp\to tth$. Thus, to rule out such a scenario conclusively, measuring the $t\bar{t}hh$ process during future LHC runs or at future colliders is not optional but a necessity.

In the present work we revisit the proposal of~\cite{Englert:2014uqa} by extending it to the potential future FCC-hh 100 TeV pp collider and including the study of contributions from effective $t\bar{t}hh$ interactions. We will focus on the scenario where both Higgs bosons decay into bottom quarks while one of the top quarks decays leptonically and the other hadronically~\footnote{A fully leptonic $t\bar{t}$ even though much cleaner, suffers from a reduction in the total rate. A fully hadronic scenario, on the other hand, will entail large QCD backgrounds. However, the total sensitivity will increase if we study the leptonic, semi-leptonic and hadronic scenarios in conjunction. This we leave for a more comprehensive future study.}. Owing to the increase in energy, we will see that this channel is competitive with various other di-Higgs channels~\cite{Yao:2013ika, Barr:2014sga, Azatov:2015oxa, He:2015spf, Contino:2016spe, Cao:2016zob, Mangano:2018mur, Benedikt:2018csr, Buchalla:2018yce, Aad:2019uzh} in constraining the trilinear Higgs self-coupling at the 100 TeV collider. This increase in cross section due to energy is of course also a feature of the backgrounds, and we therefore substantially increase their discussion.

Using the formalism of effective field theories, with a strongly-coupled UV completion in mind, in Sec.~\ref{sec:eft} we describe why obtaining a direct measurement of the $t\bar{t}hh$ final state at current or future colliders is of importance to obtain meaningful constraints in the top-Higgs sector. In Sec.~\ref{sec:mc} we describe the technical framework used. The analysis steps, reconstruction efficiencies and kinematic features of the signal and the background are detailed in Sec.~\ref{sec:analysis}. Finally, we offer our conclusions in Sec.~\ref{sec:summary}.

\section{Effective Field Theory formalism}
\label{sec:eft}

\begin{figure}[ht]
\centering
\includegraphics[scale=0.45]{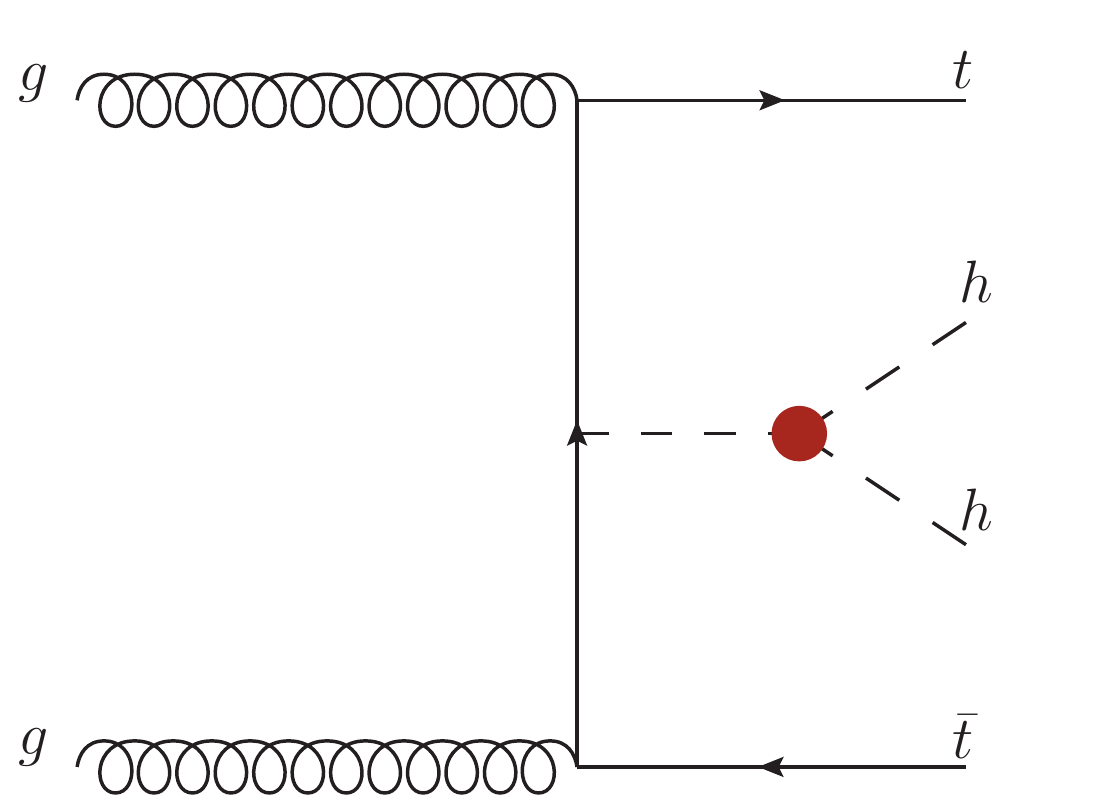}~~~\includegraphics[scale=0.45]{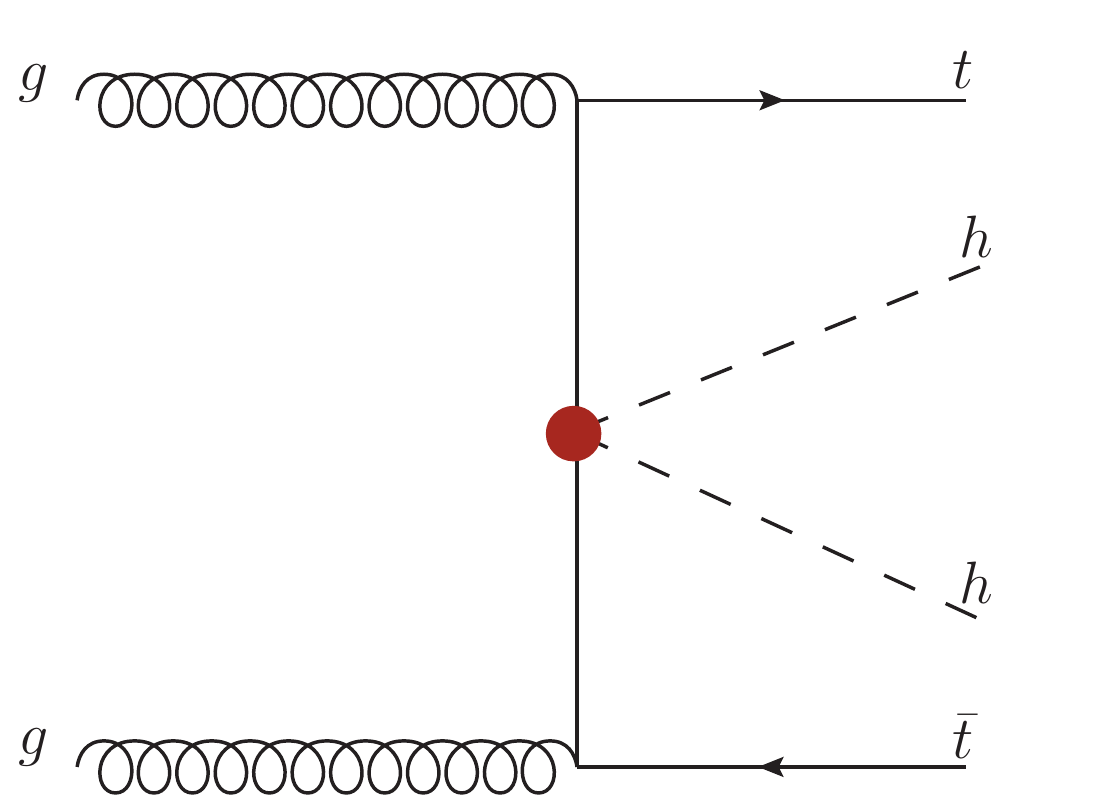}~~~\includegraphics[scale=0.45]{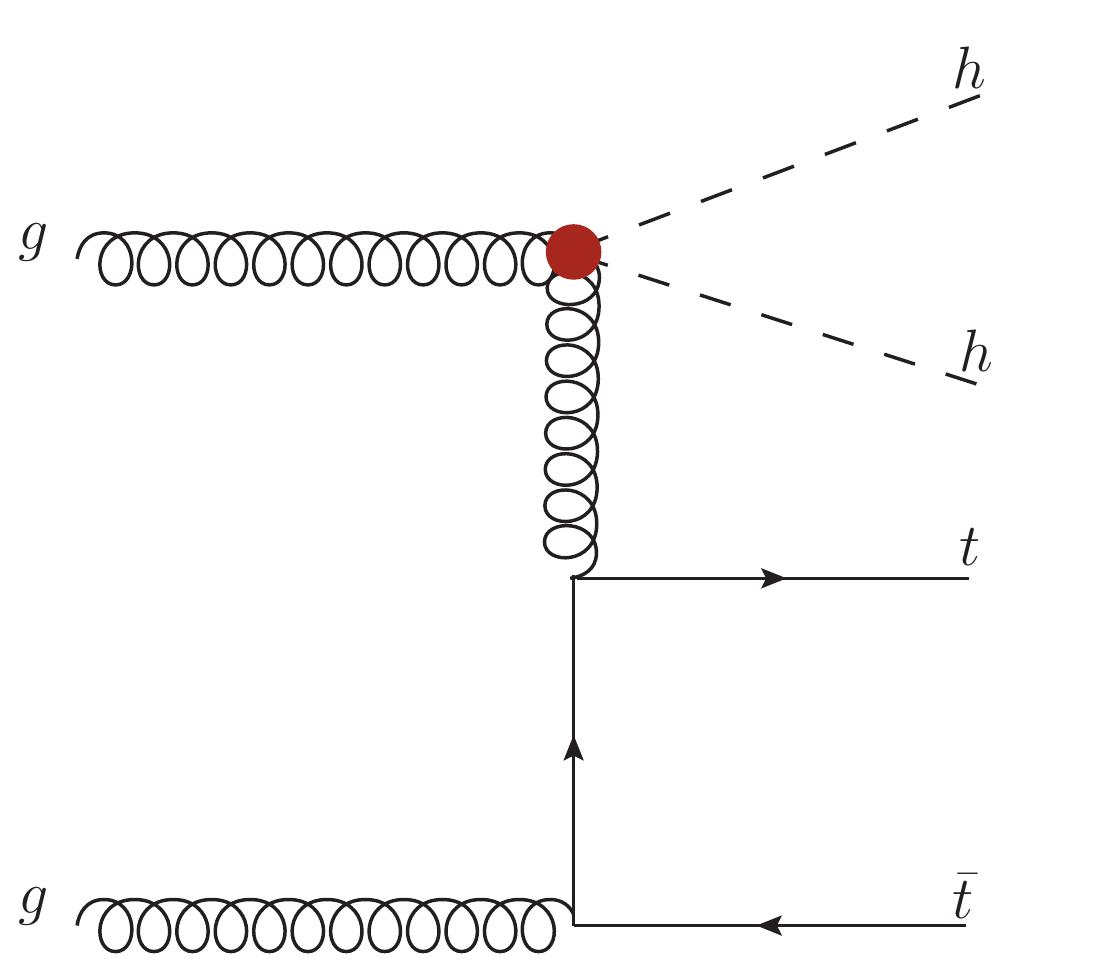}
\caption{Feynman diagrams~\cite{Binosi:2008ig} showing the impact of the three effective vertices, \textit{viz.}, $hhh, t\bar{t}hh$ and $gghh$.}
\label{fig:Feynman}
\end{figure}

Extensions of the Standard Model can lead to various modifications of Higgs interactions. Some of the most popular are composite Higgs models, which assume that the Higgs boson is a pseudo-Nambu-Goldstone boson of a strongly coupled UV completion. The most general effective field theory that describes the low-energy effects of a strongly-coupled embedding of the Standard Model is the electroweak chiral Lagrangian (ew$\chi$L)~\cite{Feruglio:1992wf,Bagger:1993zf,Koulovassilopoulos:1993pw,Buchalla:2012qq,Buchalla:2013rka}. Here, the $SU(2) \times U(1)$ symmetry is non-linearly realised,
\begin{equation}
\Sigma(x) = e^{i \sigma^a \phi^a(x)/v},
\end{equation}
with the Goldstone bosons $\phi^a$ (a=1,2,3) and the Pauli matrices $\sigma^a$. After introducing a scalar field that transforms linearly under the custodial symmetry, the Lagrangian contains\footnote{We follow the notation of \cite{Gillioz:2012se}. In~\cite{Gillioz:2012se} it is also shown how the coefficients $k_g, k_{2g},c,c_2$ and $d_3$ translate to the effective dimension-6 operators of a linearised sigma model, the so-called SILH parametrisation~\cite{Giudice:2007fh}.} 
\begin{eqnarray}
\label{eq:lag}
\mathcal{L}^{\mathrm{ew} \chi } \supset & - & V(h) + \frac{g_s^2}{48 \pi^2} G_{\mu\nu}^a G^{\mu\nu}_a \left( k_g \frac{h}{v} + \frac{1}{2} k_{2g} \frac{h^2}{v^2} + \cdots \right ) \\ \nonumber
& - & \frac{v}{\sqrt{2}} (\bar{u}_L^i~\bar{d}^i_L) \Sigma \left [ 1+ c \frac{h}{v} + c_2 \frac{h^2}{v^2} + \cdots \right ] \begin{pmatrix} y^u_{ij} u_R^j \\ y^d_{ij} d_R^j  \end{pmatrix} + \hc ,
\end{eqnarray}
with 
\begin{equation}
\label{eq:V}
V(h) = \frac{1}{2} m_h^2 h^2 + d_3 \frac{m_h^2}{2v} h^3 + d_4 \frac{m_h^2}{8 v^2} h^4 + \cdots~.
\end{equation}
Focusing on contributions of effective operators to the top-Higgs sector we find 5 operators to be of imminent importance, \textit{i.e.}, the ones associated with the coefficients $k_g$, $k_{2g}$, $c$, $c_2$ and $d_3$. While $k_g$ and $c$ can be constrained in various single-Higgs production processes, \textit{e.g.} gluon-fusion, vector-boson fusion or top-associated single-Higgs production, the coefficients $k_{2g}, c_2$ and $d_3$ rely at leading-order predominantly on double-Higgs production processes to be tensioned with data. Thus, to over-constrain the parameter space of $\mathcal{L}^{\mathrm{ew}\chi}$ it is necessary to access as many double-Higgs channels as possible, \textit{i.e.} $pp \to hh$, $pp \to hhjj$ and $pp \to t\bar{t}hh$. The process $pp \to t\bar{t}hh$ is of particular relevance to constrain $c_2$, as it is the only process of appreciable cross section where this coefficient can be constrained at tree-level. However, it is to be noted that the one-loop gluon fusion production of di-Higgs (at LO) also affects $c_2$, albeit with a different weight from $t\bar{t}hh$. The Feynman diagrams showing the modified vertices are shown in Fig.~\ref{fig:Feynman}.

In this paper, we work with the simplified Lagrangian
\begin{equation}
\label{eq:Lsimp}
\mathcal{L}^{\mathrm{simp}} = \mathcal{L}^{SM} + (1 - \kappa_{\lambda})\lambda_{\textrm{SM}} h^3 + \kappa_{t\bar{t}hh} (\bar{t}_L t_R h^2 + \hc) - \frac{1}{8} \kappa_{gghh} G_{\mu\nu}^a G^{\mu\nu}_a h^2,
\end{equation}
where $\lambda_{\textrm{SM}} = \lambda v = \frac{m_h^2}{2v}$~\footnote{The Higgs potential in the SM can be written as $V_H = \frac{1}{2}{2 \lambda v^2}h^2 + \lambda v h^3 + \frac{\lambda}{4} h^4$.} and $\kappa_{\lambda} = \lambda_{\textrm{BSM}}/\lambda_{\textrm{SM}}$. In Table~\ref{tab:map}, we show the relations between the various bases. For reference, we also include the relationship with the SILH basis~\cite{Alloul:2013naa, Giudice:2007fh, Gillioz:2012se,Goertz:2014qta}, which corresponds to a linearised sigma model. The SILH Lagrangian is defined as~\cite{Gillioz:2012se}

\begin{align}
\mathcal{L}_{\textrm{SILH}} = \frac{c_H}{2f^2}\partial^{\mu} (H^{\dagger}H) \partial_{\mu}(H^{\dagger}H) + \frac{c_{\tau}}{2f^2} H^{\dagger}H (D^{\mu}H)^{\dagger} (D_{\mu}H) - \frac{c_6 \lambda}{f^2} (H^{\dagger}H)^3 \nonumber \\
+ (\frac{c_y y_f}{f^2} H^{\dagger}H \bar{f}_L H f_R + \textrm{h.c.}) + \frac{c_g g_s^2}{16 \pi^2 f^2} \frac{y_t^2}{g_{\rho}^2} H^{\dagger}H G^a_{\mu\nu} G^{a \mu \nu} + \frac{c_{\gamma} {g'}^2}{16 \pi^2 f^2} \frac{g^2}{g_{\rho}^2} H^{\dagger}H B_{\mu\nu} B^{\mu \nu},
\end{align}
where $g, g_s$ and $g'$ are respectively the $SU(2)_L, SU(3)_c$ and $U(1)_Y$ couplings in the SM, $g_{\rho}$ is the coupling of the strongly interacting sector, and $\lambda$ and $y_f$ are respectively the Higgs quartic coupling and the Yukawa coupling.
\begin{table}
  \begin{center}
    \begin{tabular}{||c|c|c|c||}
      \hline
   Coupling     & Non-linear EFT                        & Simplified Lagrangian & SILH \\
   \hline\hline
   $\displaystyle{\vphantom{\frac{|^|_|}{|^|_|}}}$
   $hhh$        & $\displaystyle{d_3}$                  & $\displaystyle{\kappa_{\lambda}}$                           & $\displaystyle{1 + (c_6 - c_{\tau}/4 - 3 c_H/2)\xi}$ \\
   $\displaystyle{\vphantom{\frac{|^|_|}{|^|_|}}}$
   $t\bar{t}hh$ & $\displaystyle{c_2}$                  & $-\displaystyle{\frac{\sqrt{2}v}{y_t}\kappa_{t\bar{t}hh}}$  & $-\displaystyle{(c_H + 3 c_y + c_{\tau}/4)\xi/2}$               \\
   $\displaystyle{\vphantom{\frac{|^|_|}{|^|_|}}}$
   $gghh$       & $\displaystyle{k_{2g}}$               & $-\displaystyle{\frac{12 \pi^2 v^2}{g_s^2}\kappa_{gghh}}$   & $\displaystyle{3 c_g\Big(\frac{y_t^2}{g_{\rho}^2}\Big)\xi}$                  \\
      \hline\hline
    \end{tabular}
    \caption{Relationship between the $hhh$, $t\bar{t}hh$ and $gghh$ vertices in three different bases~\cite{Alloul:2013naa, Giudice:2007fh, Gillioz:2012se}, where $\xi \equiv (v/f)^2$.}
    \label{tab:map}
  \end{center}
\end{table}

\begin{figure}[ht]
\centering
\includegraphics[scale=0.5]{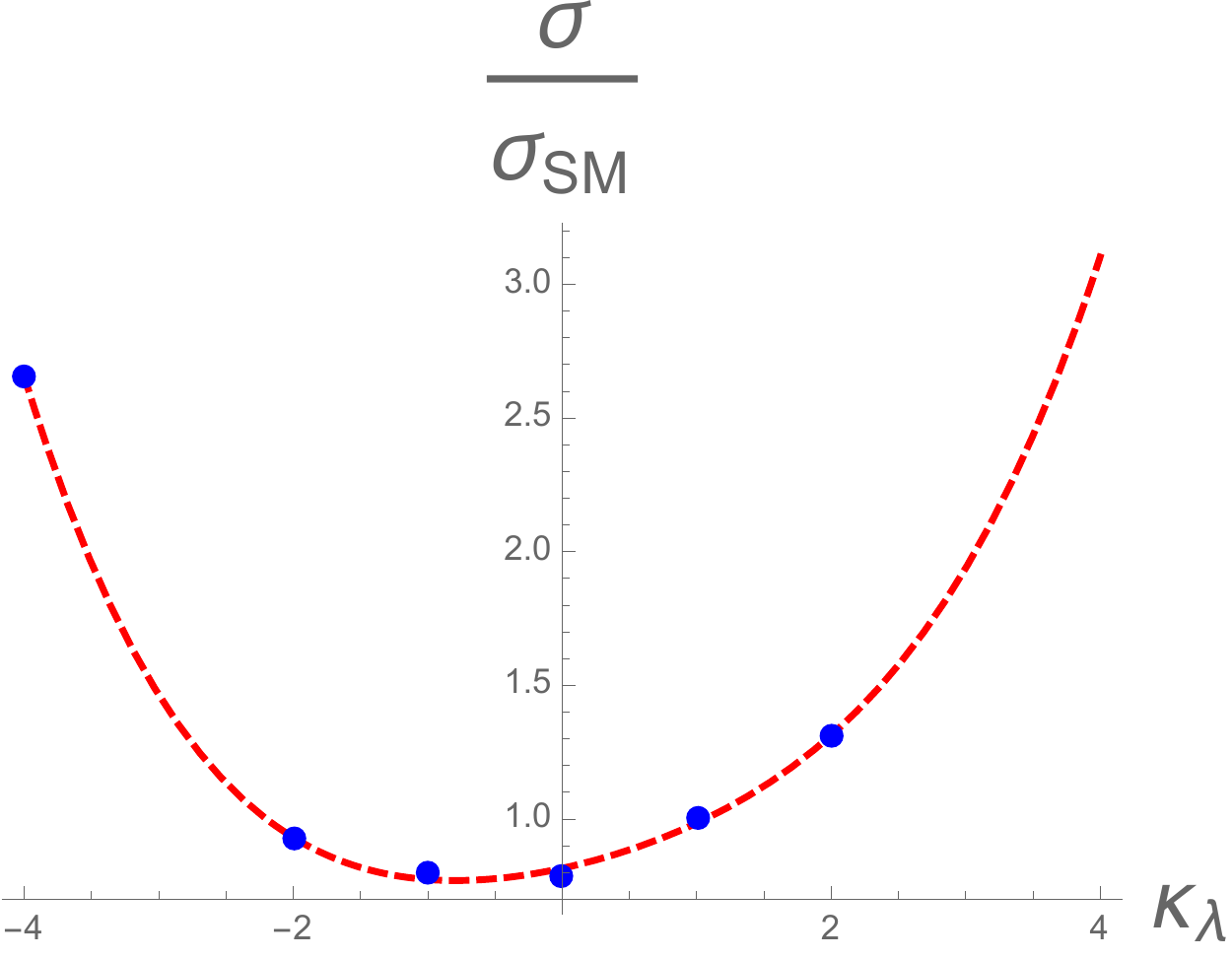}~~~~\includegraphics[scale=0.5]{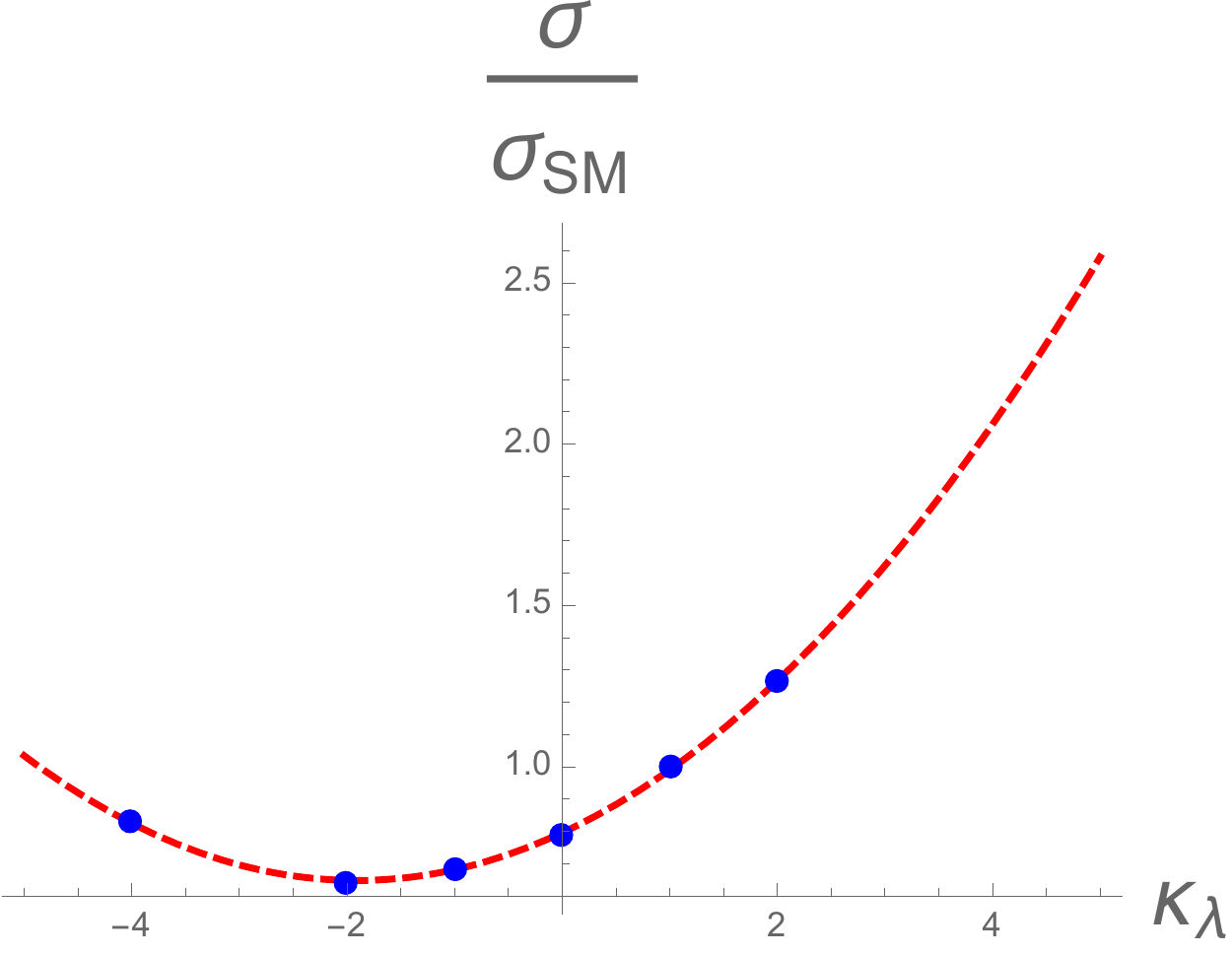}\\
\includegraphics[scale=0.4]{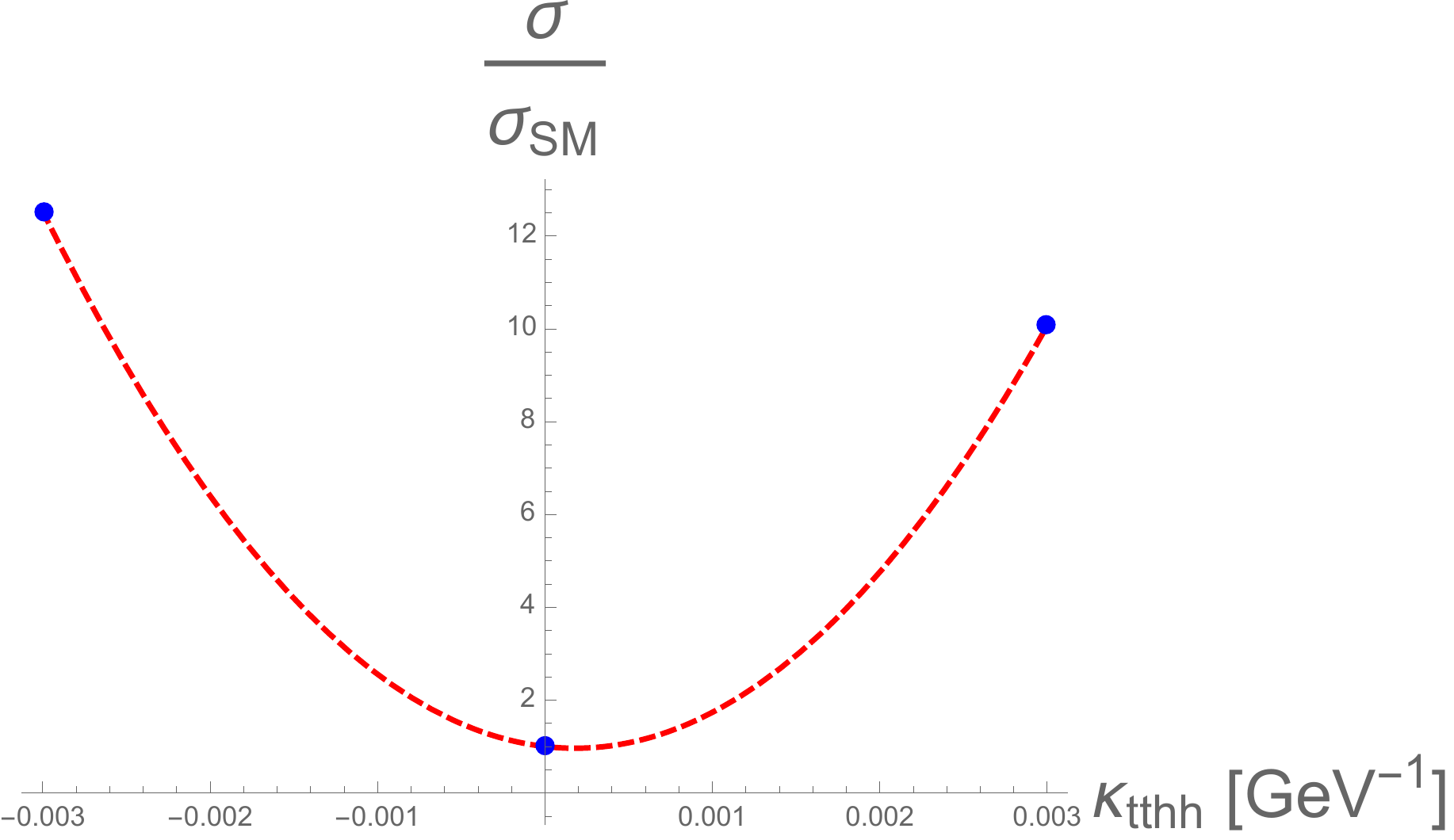}~~~~\includegraphics[scale=0.4]{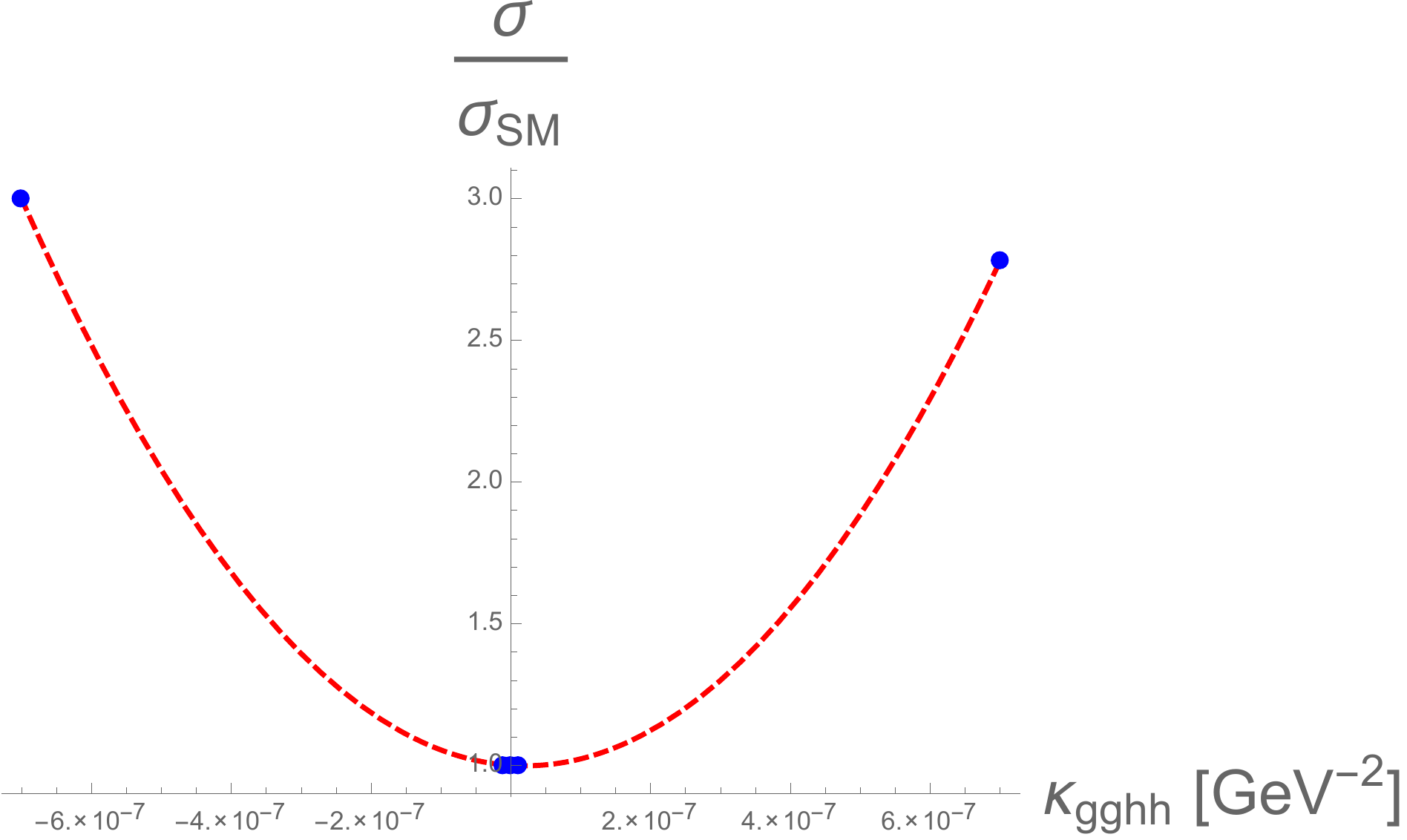}
\caption{$\sigma/\sigma_{\textrm{SM}}$ as a function of $\kappa_{\lambda}$ (top left at 100 TeV and top right at 14 TeV), $\kappa_{t\bar{t}hh}$ [GeV$^{-1}$] (bottom left at 100 TeV) and $\kappa_{gghh}$ [GeV$^{-2}$] (bottom right at 100 TeV). For the 100 TeV case, $\sigma_{\textrm{SM}} = 16.4$ fb (this includes $h \to b\bar{b}$ for both Higgs bosons). The corresponding cross-section for the 14 TeV scenario is 0.22 fb.}
\label{fig:signalXS}
\end{figure}

We show the ratio of signal cross-sections with respect to the SM expectation as a function of $\kappa_{\lambda}$, $\kappa_{t\bar{t}hh}$ and $\kappa_{gghh}$ in Fig.~\ref{fig:signalXS}. One can see that the cross-section increases with $\lambda > \lambda_{\textrm{SM}}$. We validated our setup by checking this ratio ($\sigma/\sigma_{\textrm{SM}}$) at 14 TeV with Ref.~\cite{Frederix:2014hta}. It is interesting to note that the nature of the growth of the cross-section for negative values of $\lambda$ changes significantly upon going from a 14 TeV machine to a 100 TeV machine. In the SILH Lagrangian, the coupling modifying the $gghh$ coupling also contributes to the $ggh$ coupling. Allowing for a 10\% modification in the 14 TeV $gg \to h$ cross-section, we find that $\kappa_{gghh}$ is very strongly constrained. Thus, in the analysis, we only consider the couplings modifying the trilinear Higgs coupling and the $t\bar{t}hh$ vertex.

Before moving on with the analysis, we dedicate this paragraph to explain the differences between the linear and the non-linear realisations of the EFT. Upon considering a linear EFT, the Higgs is essentially considered to be the part of a doublet (as in SM) and the couplings affecting the $ggh$ and the $gghh$ vertices are coming from the same set of Wilson coefficients. This is the same for the $t\bar{t}h$ and the $t\bar{t}hh$ couplings and also for the trilinear and the quartic Higgs couplings. Hence, in a linear realisation of the EFT, we have less number of parameters to constrain~\footnote{This statement is true provided we have a fixed order in mass dimension.}. On the contrary, when one considers a non-linear realisation of the EFT, as we will be considering for most of this work, all these couplings are independent and a priori there are no preferential directions. In Ref.~\cite{Azatov:2015oxa}, the $p p \to h h$ process has been studied in the non-linear EFT scenario where the various parameters have been constrained upon studying the $hh \to b\bar{b}\gamma\gamma$ channel. From Ref.~\cite{Azatov:2015oxa}, it can be seen that $\kappa_{gghh}$ is constrained to around $\mathcal{O}(10^{-8})$ GeV$^{-2}$ from the $pp \to hh$ production, which is an order of magnitude stronger than our assumption~\footnote{The 68\% constraints derived from Ref.~\cite{Azatov:2015oxa} are $\kappa_{gghh} \in [-1.73, 4.97] \times 10^{-8}$ GeV$^{-2}$, for an integrated luminosity of 3/ab, while considering $d_3 = 1$ and $\kappa_g = 0$, the SM values.}. It is important to realise that various double Higgs processes give us constraints on different linear combinations of these couplings and one encounters blind directions. This necessitates the study of all double Higgs processes in order to break these blind directions and to combine the results to obtain stronger constraints. In this first work for the $p p \to t\bar{t}hh$ channel, we consider the non-linear realisation of the EFT and treat the couplings independent of each other.

\section{The Monte Carlo Setup}
\label{sec:mc}

The final state in our analysis results from the decays of the two top quarks and the two Higgs bosons into six $b$-tagged jets, one isolated lepton, missing transverse momentum, and at least two extra jets which are not $b$-tagged.  This leaves a wide range of backgrounds to be considered, see below.  In all channels, potential additional jets may give rise to required final state particles, either by jet radiation mimicking the light jets stemming from the hadronically decaying top quark, or by gluons splitting into $b$-quark pairs, yielding $b$-tagged jets.  In addition, light and charm jets may produce fake $b$ tags. Apart from our signal $t\bar{t}HH$ we therefore include as backgrounds processes where
\begin{itemize}
\item $Z$ and Higgs bosons decay into $b$-quark pairs, such as $t\bar{t}ZZ$ and $t\bar{t}hZ$;
\item one or both pairs of $b$-quarks are produced through gluon splittings in QCD, like
  $t\bar{t} h b\bar{b}$, $t\bar{t}Z b\bar{b}$, or $t\bar{t}b\bar{b}b\bar{b}$; 
\item the leptonically decaying top quark is mimicked by a $W$ plus four $b$-jets,
  such as $W^{\pm} b\bar{b} b\bar{b}+$jets; and
\item sub-dominant or fake backgrounds which can contribute to the total background yield,
  for example $t\bar{t}c\bar{c}c\bar{c}$ and $W^{\pm}c\bar{c}c\bar{c}$, misidentifying charm
  as $b$-quarks, or $t\bar{t}b\bar{b}$, $t\bar{t}h$, $t\bar{t}Z$, and $W^{\pm}b\bar{b}$,
  all associated by light jets.
\end{itemize}
Due to their complexity and their large final-state multiplicity we chose to
simulate signal and backgrounds with leading order matrix elements and consistently
combine them with subsequent parton showers, to capture the effect of QCD radiation and,
in particular, of gluon splittings into heavy quarks.

We use \texttt{SHERPA} v2.2.5~\cite{Gleisberg:2003xi,Gleisberg:2008ta} with the \texttt{COMIX}
matrix element generator~\cite{Gleisberg:2008fv} and the parton shower based on Catani-Seymour
splitting kernels~\cite{Nagy:2007ty,Schumann:2007mg}. The central \texttt{CT14NLO} PDF
set~\cite{Dulat:2015mca} is used throughout. All jets contributing to the process classification 
including $b$ and $c$-jets, are defined with the anti-$k_T$ algorithm~\cite{Cacciari:2008gp} with
\begin{equation}
  p_T(j) >25\;\mbox{\rm GeV}\;,\;\;\;|y_j|<4.0.
\end{equation}
We also require a minimal invariant mass $m_{bb/bc/cc} \geq 50$ GeV for all possible $b$ and $c$ pairs.
Where necessary, we add matrix elements for final state with more jets through multijet merging
according to~\cite{Catani:2001cc,Krauss:2002up,Hoeche:2009rj}, and a merging cut of
$Q_{\rm cut} = 40$ GeV.
For the various processes we use the renormalisation and factorisation
scales listed in Table~\ref{tab:scales}, where for merged samples we cluster back to the relevant
core process before determining the scales.

\begin{table}
  \begin{center}
    \begin{tabular}{|c|c|c|}
      \hline
      Process category & $\mu^2_{F}$ & $\mu^2_{R}$\\
      \hline\hline
      $t\bar{t}HH$, $t\bar{t}ZZ$, $t\bar{t}HZ$ & $\frac14\,H_T^2+2m_t^2+\{2m_H^2, 2m_Z^2, m_H^2+m_Z^2\}$
                                               & $\frac14\,H_T^2+2m_t^2$\\
      $t\bar{t}Hb\bar{b}$, $t\bar{t}Zb\bar{b}$ & $\frac14\,H_T^2+m^2_{H,Z}+2m_t^2$
                                               & $\frac14\,H_T^2+2m_t^2$\\
      $t\bar{t}$ + $b$'s, $c$'s or light jets  & $\frac14\,H_T^2+2m_t^2$
                                               & $\frac14\,H_T^2+2m_t^2$\\
      $W$ + $b$'s, $c$'s or light jets         & $\frac14\,H_T^2+m_W^2$
                                               & $\frac14\,H_T^2$\\
      \hline
    \end{tabular}
    \caption{Renormalisation and factorisation scales used for the various processes, where $H_T$ is the scalar sum of the $p_T$ of the final state particles.}
    \label{tab:scales}
  \end{center}
\end{table}

Details of the generation cross-sections for all signal and background samples are listed in
Tab.~\ref{tab:xsecBeforeCuts}. 

\begin{table}[t]
\begin{center}
\begin{tabular}{||c|c||}
\hline
Channel                                       & Cross-section [pb]  \\
\hline
\hline
$t\bar{t}hh$ ($\kappa_{\lambda} = 1$)         & 0.015               \\
$t\bar{t}hh$ ($\kappa_{\lambda} = 2$)         & 0.020               \\
$t\bar{t}hh$ ($\kappa_{\lambda} = 0$)         & 0.012               \\
$t\bar{t}hh$ ($\kappa_{\lambda} = -1$)        & 0.012               \\
$t\bar{t}hh$ ($\kappa_{\lambda} = -2$)        & 0.014               \\
\hline
$t\bar{t}hh$ ($\kappa_{t\bar{t}hh} = -0.003$ GeV$^{-1}$) & 0.175               \\
$t\bar{t}hh$ ($\kappa_{t\bar{t}hh} = 0.003$ GeV$^{-1}$)  & 0.132               \\
\hline
$t\bar{t}b\bar{b}b\bar{b}$                    & 0.174               \\
$t\bar{t}c\bar{c}c\bar{c}$                    & 0.174               \\
$t\bar{t}b\bar{b}+$jets                       & 46.30               \\
$t\bar{t}h b\bar{b}$          		      & 0.076               \\
$t\bar{t}h+$jets            		      & 12.825              \\
$t\bar{t}h Z$              	              & 0.045               \\
$t\bar{t}Z Z$                		      & 0.057               \\
$t\bar{t}Z b\bar{b}$          		      & 0.165               \\
$t\bar{t}Z+$jets             		      & 25.663              \\
$W^{\pm}b\bar{b}b\bar{b}+$ jet		      & 0.036               \\
$W^{\pm}c\bar{c}c\bar{c}+$ jet		      & 0.092               \\
\hline
\hline
\end{tabular}
\caption{Table shows the generation level cross-sections for the signal and background processes. We require the Higgs bosons to decay to a pair of $b/c$ quarks, the $Z$-bosons to all quarks. Furthermore, we require the $W^{\pm}$-bosons to decay leptonically. These branching ratios are included in these cross-sections. For the signals, $\kappa_{\lambda}$, is the ratio of the Higgs self-coupling to the SM value and $\kappa_{t\bar{t}hh}$ is the coupling of the four point $t\bar{t}hh$ interaction. The processes with $b/c$ quarks in the final state in the matrix element level have a further requirement of $m_{bb/cc/bc} > 50$ GeV, $p_T(b/c) > 25$ GeV, $D$-parameter $> 0.4$, $|y| < 4.0$.}
\label{tab:xsecBeforeCuts}
\end{center}
\end{table}

\section{Analysis}
\label{sec:analysis}

In Ref.~\cite{Englert:2014uqa}, the $t\bar{t}hh$ channel was studied in the context of the 14 TeV high-luminosity run of the LHC. In the present work we revisit this analysis by focussing on the prospects of observing this channel at a possible future FCC-hh 100 TeV $pp$ collider. While our analysis strategy here is reminiscent of Ref.~\cite{Englert:2014uqa}, we study the backgrounds in significantly greater detail. 

\begin{figure}[ht]
\centering
\includegraphics[scale=0.6]{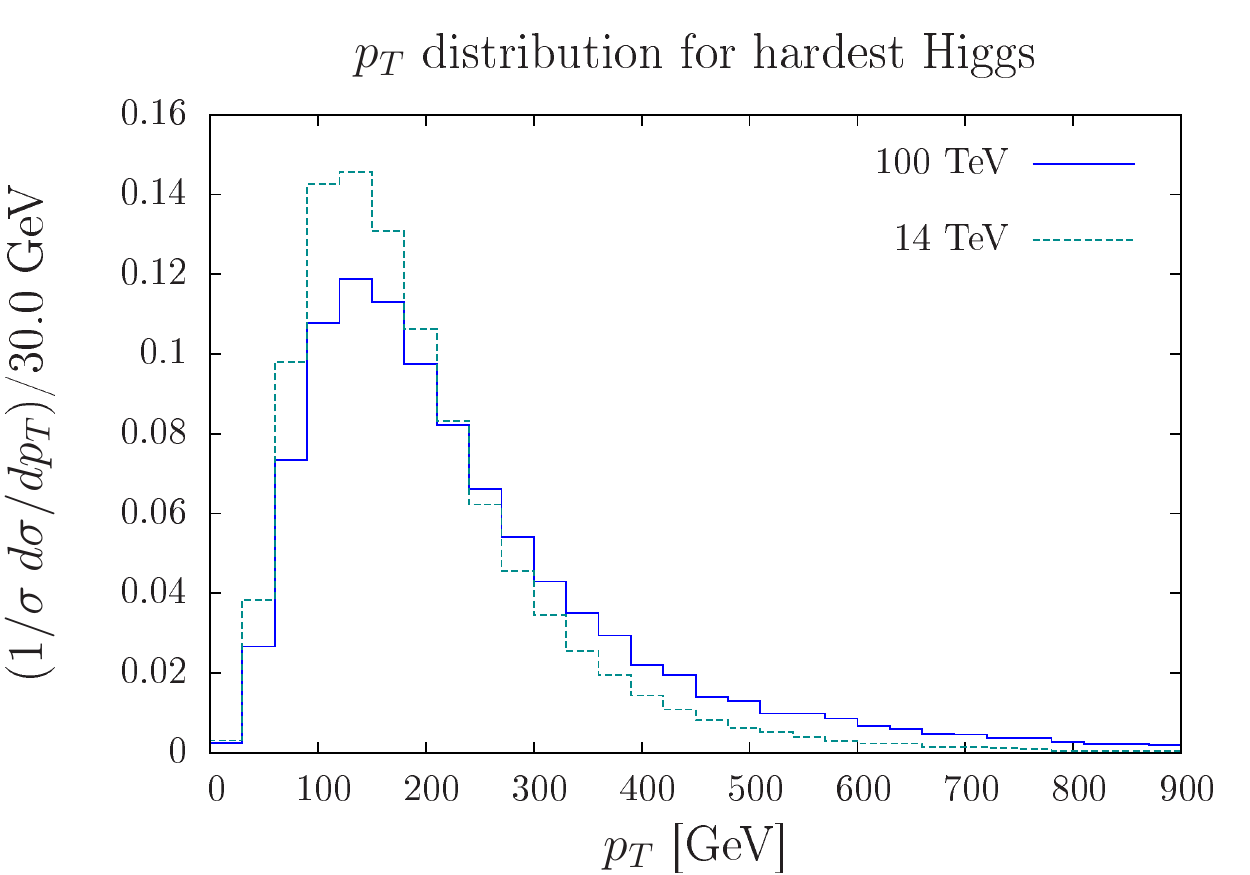}~~\includegraphics[scale=0.6]{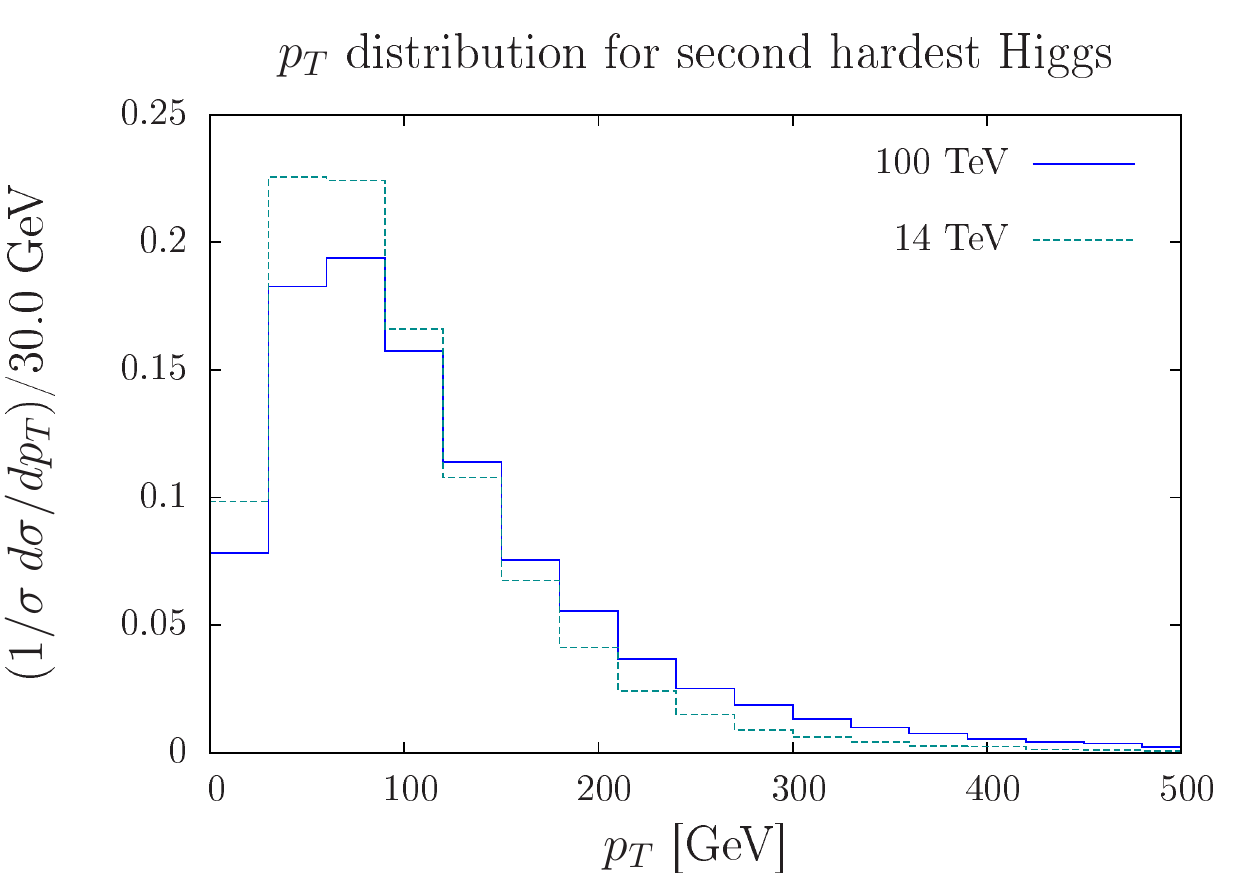}\\
\includegraphics[scale=0.6]{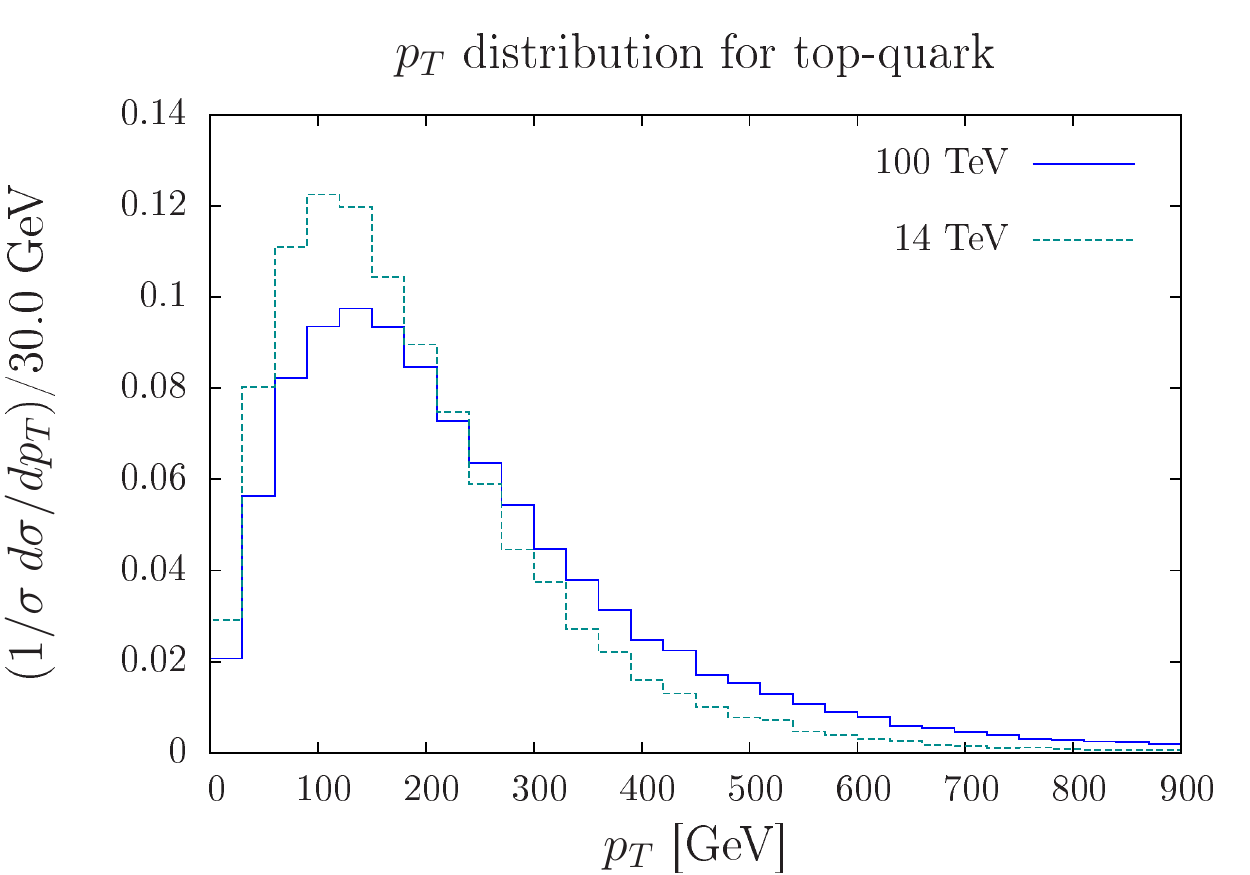}~~\includegraphics[scale=0.6]{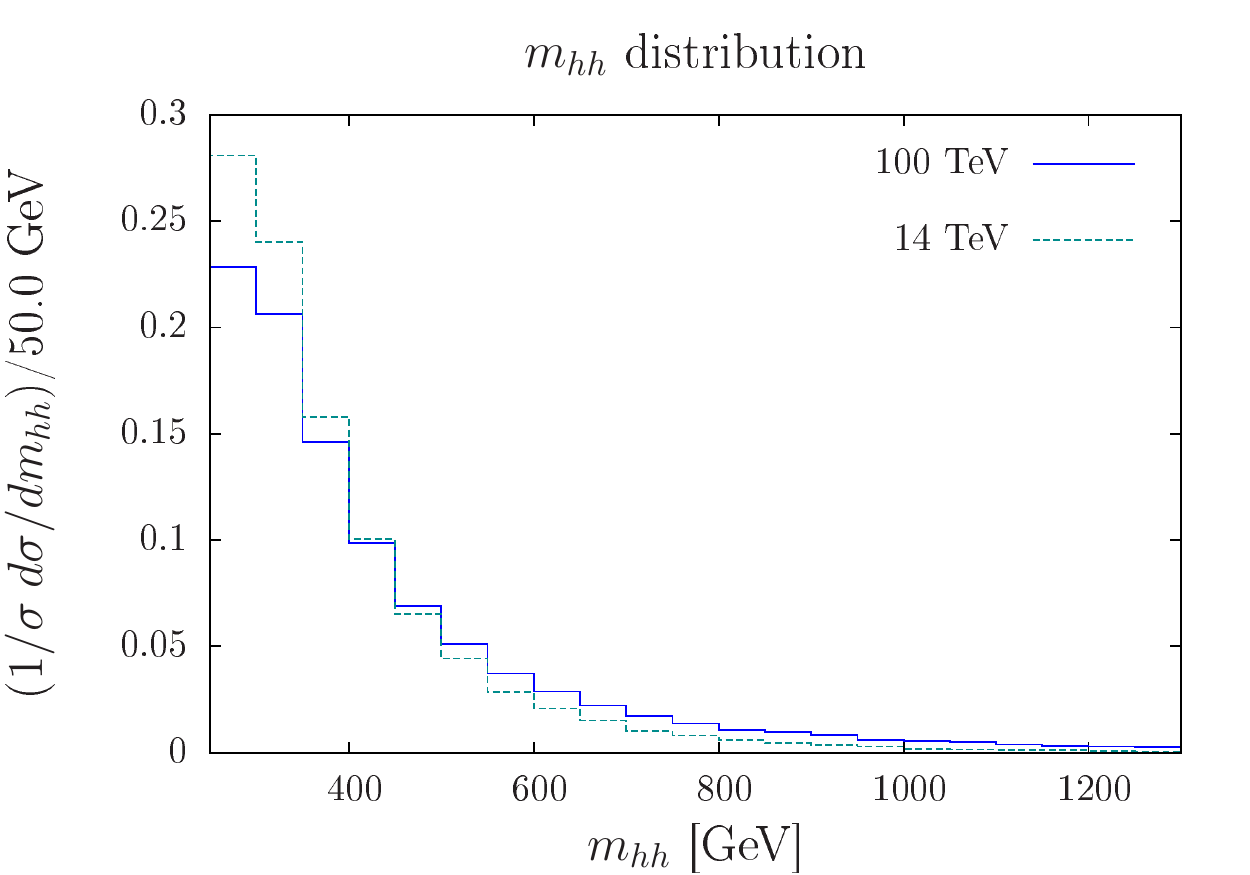}
\caption{Normalised distributions showing the $p_T$ spectra of the hardest and second hardest Higgs bosons and the top quark, and the invariant mass spectrum of the di-Higgs system, at the 14 TeV and 100 TeV colliders.}
\label{fig:pT}
\end{figure}

At leading order, the $pp \to t\bar{t} h h$ cross-section increases by a factor $\sim 75$ for the SM scenario upon going from 14 TeV to 100 TeV. However, from Fig.~\ref{fig:pT}, we can see that the large increment in the total cross-section does not translate into a significantly enhanced distribution for large transverse momenta. Hence, any advantage in the analysis is due to the increased total rate and to an improved reconstruction efficiency for highly energetic final-state objects.

To reconstruct the final state, jets are clustered with the anti-$k_t$ algorithm~\cite{Cacciari:2008gp} in the \texttt{FastJet} framework~\cite{Cacciari:2011ma} and parameters
\begin{equation}
  R=0.4\;,\;\;\;p_{T,j} > 30\,\mbox{\rm GeV}\;,\;\;\;\mbox{\rm and}\;\;\;|\eta_j| < 4.5\,.
\end{equation}
We require our events to have at least 8 jets, out of which exactly 6 must be $b$-tagged.  For the $b$-tagged jets we demand that the distance between $B$-hadron and jet center fulfils
\begin{equation}
  \Delta R_{j,B}<0.2
\end{equation}
and that
\begin{equation}
  |y_b| < 2.5\,.
\end{equation}
For the 100 TeV collider study, we assume a $b$-tagging efficiency of 80\%. To estimate the effect of fake $b$-tags, we assume a mistagging probability of 10\% for $c$-jets and of 1\% for light jets. We require exactly one lepton in each event with
\begin{equation}
  p_T(\ell) > 10\mbox{\rm GeV}\;\;\;\mbox{\rm and}\;\;\;|y_{\ell}| < 2.5\,.
\end{equation}
To isolate the leptons we demand that the total hadronic activity around a cone of $\Delta R = 0.3$ to be less than 10\% of its $p_T$.

From Tab.~\ref{tab:xsecBeforeCuts}, it is clear that some of the backgrounds are much larger in cross-sections than others. We note that the $t\bar{t}h+$jets process already contains $t\bar{t}h b\bar{b}$. The same is true for $t\bar{t}Z+$jets and $t\bar{t}Zb\bar{b}$. Thus, in order to avoid double counting, we focus on the $t\bar{t}h+$jets and $t\bar{t}+$jets channels. For the $t\bar{t}h/Z+$jets, we consider the merged sample, where additional jets stem from QCD radiation, including the gluon splitting into $b$-quark pairs. In order to ensure that none of the additional jets contains more than one $B$-mesons, we implement a further criterion, namely that the $B$-hadron closest to the jet axis satisfies
\begin{equation}
  x_B\,=\,\frac{|\vec{p}_{B}|}{|\vec{p}_j|}\times
  \frac{\vec{p}_{B} \cdot \vec{p}_j}{|\vec{p}_{B}| |\vec{p}_{j}|} > 0.7\,.
\end{equation}
This condition reflects the $b$-quark fragmentation, which is characterised by relatively low energy or momentum losses due to QCD radiation and, correspondingly, the fact that $B$-hadrons in a jet stemming from a $b$-quark carry the dominant fraction of the jet momentum.  Obviously, this is not true for those jets, where a gluon splits into two $b$-quarks, which typically have relatively symmetric momentum fractions.  Consequently, this criterion effectively suppresses ``doubly-tagged'' $b$-jets, which contain two $b$-hadrons.  $b$-tagged jets failing to fulfil this criterion are considered as light jets, and we call those jets that fulfil it ``good'' $b$-jets.  We therefore require events with exactly 6 \textit{good} $b$-tagged jets and all pairs of $b$-tagged jets must have an invariant mass greater than 50 GeV. We confirmed the results of some previous investigations using such a cut, for example in~\cite{Goncalves:2015prv}, which found that it suppresses each ``doubly-tagged'' $b$-jets by more than about 80\%.  As a consequence, we could confirm that these additional conditions on ``good'' $b$-jets render the effect of gluon jets tagged as $b$-jets due to gluon splitting negligibly small.

\begin{figure}[ht]
\centering
\includegraphics[scale=0.6]{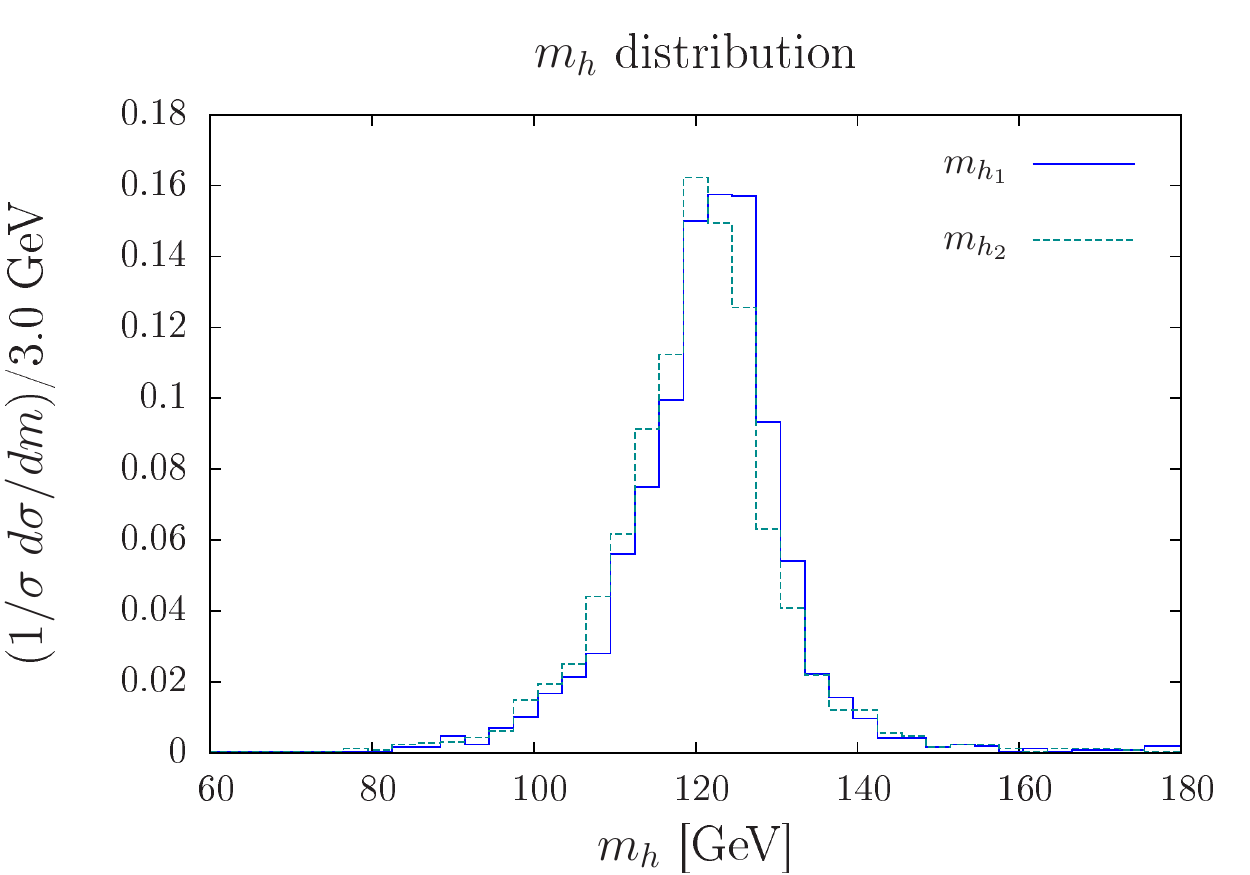}~~\includegraphics[scale=0.6]{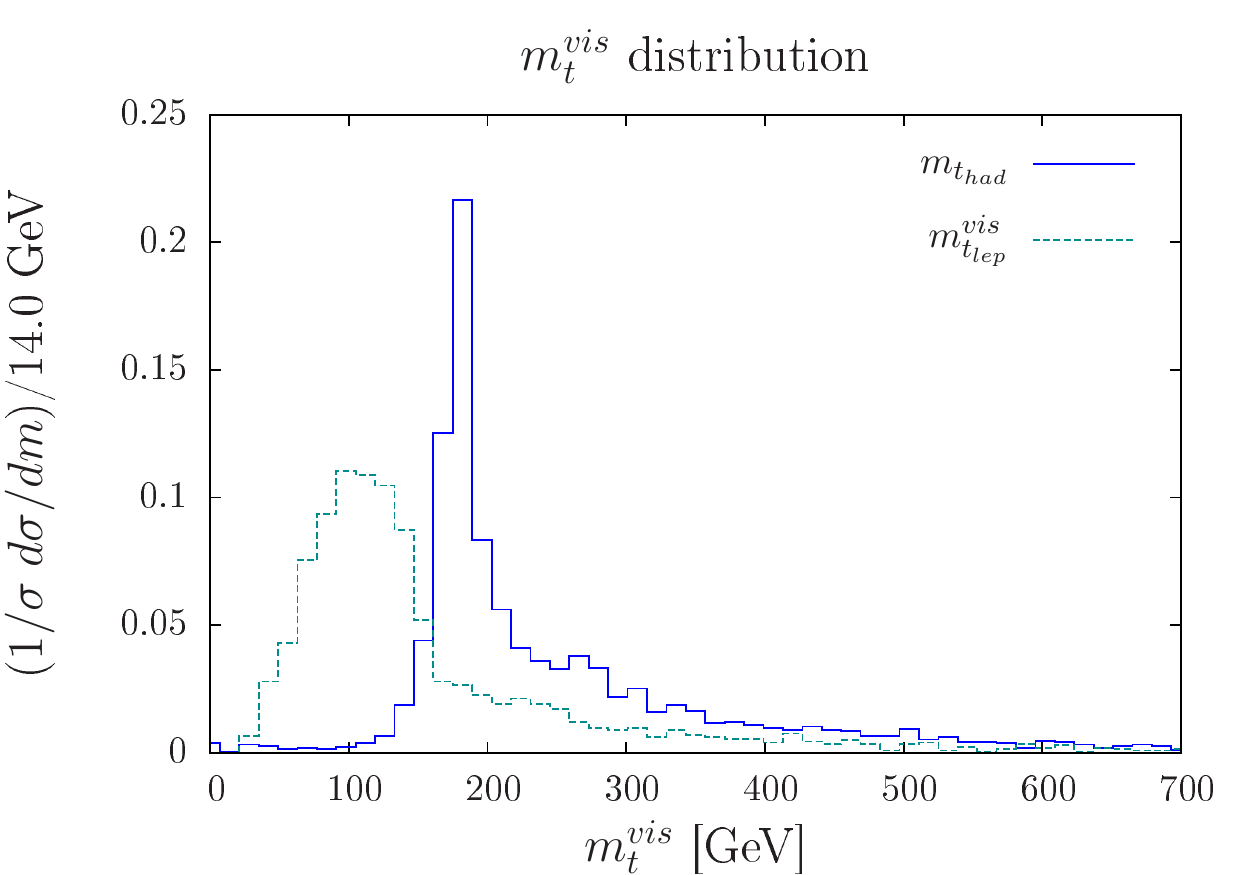}
\caption{Mass reconstruction of the hardest and the second-hardest Higgs bosons (left) and the hadronic and visible part of the leptonic top quarks (right).}
\label{fig:HiggsTop}
\end{figure}

In order to ascertain that the events are ensuing from a $t\bar{t}hh$ topology, reconstructing most of the electroweak resonances is of essence. We follow Ref.~\cite{Englert:2014uqa} and define our two Higgs boson candidates by minimizing
\begin{equation}
\chi_{HH}^2 = \frac{(m_{b_i,b_j} - m_h)^2}{\Delta_h^2} + \frac{(m_{b_k,b_l} - m_h)^2}{\Delta_h^2},
\end{equation}
where $i \neq j \neq k \neq l$ run over all the 6 $b$-tagged jets. As parameters for this minimisation we use $m_h = 120$ GeV and $\Delta_h = 20$ GeV. The strange choice for $m_h$ warrants an elucidation. Because the Higgs bosons decay to $b$-quarks which essentially hadronise to $B$-mesons, the invisible decays of the latter shifts the reconstruction of the Higgs peak to smaller values. A different value of $m_h$ can be chosen after correcting explicitly for jet energy effects in b-jets. After minimising $\chi_{HH}^2$, we require $|m_{b_i,b_j} - m_h| < \Delta_h$ and $|m_{b_k,b_l} - m_h| < \Delta_h$. The reconstructions of the hardest and the second hardest Higgs bosons are shown in Fig.~\ref{fig:HiggsTop} (left).

After finding the two Higgs bosons, we are left with two $b$-tagged jets. Because of the uncertainty in the longitudinal momentum $p_z$ of the neutrino, we only reconstruct the hadronic top, $t_h$. We consider the two remaining $b$-tagged jets and all other light jets to minimise the following $\chi^2$.
\begin{equation}
\chi_{t_h}^2 = \frac{(m_{b_i,j_k,j_l} - m_t)^2}{\Delta_t^2},
\end{equation}
where $i$ runs over the remaining two $b$-tagged jets and $k \neq l$ denote the indices of all the light jets. For this minimisation, we take $m_t = 170$ GeV and $\Delta_t = 40$ GeV. We allow for a larger uncertainty as we demand the hadronic top to be reconstructed from three jets. After minimising $\chi_{t_h}^2$, we require $|m_{b_i,j_k,j_l} - m_t| < \Delta_t$. Finally, instead of fully reconstructing the leptonic top, we reconstruct the invariant mass of the last $b$-jet which is neither part of the two Higgs reconstructions nor of the hadronic top and the single isolated lepton. We further require the visible invariant mass to satisfy $m_{t_{\textrm{lep}}^{\textrm{vis}}} < m_t$. The reconstruction of the hadronic and the visible part of the leptonic top masses (before imposing the cuts) are shown in Fig.~\ref{fig:HiggsTop} (right). These two reconstructions ensure the complete obliteration of the $W^{\pm}+$jets backgrounds. In Tab.~\ref{tab:cutflow}, we show the effects of the various cuts for three signal scenarios ($\kappa_{\lambda} = 1,2$ and $\kappa_{t\bar{t}hh} = -0.003$ GeV$^{-1}$) and the four dominant backgrounds, \textit{i.e.} $t\bar{t}b\bar{b}b\bar{b}, t\bar{t} h b\bar{b}, t\bar{t}Zh$ and $t\bar{t}Zb\bar{b}$.

\begin{table}[t]
\begin{center}
\footnotesize
\begin{tabular}{||c|c|c|c|c|c|c|c||}
\hline
Cuts                                            & $\kappa_{\lambda}=1$ & $\kappa_{\lambda}=2$ & $\kappa_{t\bar{t}hh}=-0.003$ & $t\bar{t}b\bar{b}b\bar{b}$ & $t\bar{t}h b\bar{b}$ & $t\bar{t}Zh$ & $t\bar{t}Zb\bar{b}$ \\
\hline
\hline
Trigger + isolation                             & 0.24                 &  0.24                & 0.23                         & 0.25                       & 0.26                & 0.25         & 0.25 \\
$> 7$ jets                                      & 0.56                 &  0.57                & 0.67                         &             0.68                       & 0.61                & 0.55         & 0.60 \\
6 \textit{good} $b$-jets                        & 0.019                & 0.020                & 0.018                        &             0.009                      & 0.014               & 0.004        & 0.003 \\
100 GeV $< m_{h_{1/2}} < 140$ GeV               & 0.88                 & 0.89                 & 0.84                         &             0.38                       & 0.58                & 0.68         & 0.37 \\
130 GeV $< m_{t_h} < 210$ GeV                   & 0.30                 & 0.28                 & 0.38                         & 0.30                       & 0.27                & 0.28         & 0.23 \\

$m_{t_{\textrm{lep}}^{\textrm{vis}}} < 170$ GeV & 0.76                 & 0.77                 & 0.73                         & 0.57                       &          0.60                &    0.68      & 0.47 \\
\hline
\hline
\end{tabular}
\normalsize
\caption{Cut-flow table showing the cut efficiencies for the various cuts used for three signal scenarios and for the four dominant backgrounds.}
\label{tab:cutflow}
\end{center}
\end{table}

\begin{table}[t]
\begin{center}
\begin{tabular}{||c|c||}
\hline
Channel                                  & Cross-section [fb]                     \\
\hline
\hline
$t\bar{t}hh$ ($\kappa_{\lambda} = 1$)   	& 0.0091                          \\
$t\bar{t}hh$ ($\kappa_{\lambda} = 2$)   	& 0.0118                          \\
$t\bar{t}hh$ ($\kappa_{\lambda} = 0$)  		& 0.0071                          \\
$t\bar{t}hh$ ($\kappa_{\lambda} = -1$)  	& 0.0072                          \\
$t\bar{t}hh$ ($\kappa_{\lambda} = -2$)          & 0.0083                          \\
\hline
$t\bar{t}hh$ ($\kappa_{t\bar{t}hh} = -0.003$ GeV$^{-1}$)   & 0.1135                          \\
$t\bar{t}hh$ ($\kappa_{t\bar{t}hh} = 0.003$ GeV$^{-1}$)    & 0.0912                          \\
\hline
$t\bar{t}b\bar{b}b\bar{b}$              	& 0.0217                          \\
$t\bar{t}c\bar{c}c\bar{c}$              	& $\lesssim \mathcal{O}(10^{-4})$ \\
$t\bar{t}h+$jets            			& 0.0333                          \\
$t\bar{t}h Z$              	        	& 0.0031                          \\
$t\bar{t}Z Z$                			& $\lesssim \mathcal{O}(10^{-4})$ \\
$t\bar{t}Z+$jets             			& 0.0035                          \\
$W^{\pm}b\bar{b}b\bar{b}+$ jet			& $\lesssim \mathcal{O}(10^{-4})$ \\
$W^{\pm}c\bar{c}c\bar{c}+$ jet			& $\lesssim \mathcal{O}(10^{-4})$ \\
$W^{\pm}b\bar{b}+$jets        			& $\lesssim \mathcal{O}(10^{-4})$ \\
\hline
Total Background                        	& 0.0623                          \\
\hline
\hline
\end{tabular}
\caption{Cross-sections (in fb) for the various signal scenarios and backgrounds after all the cuts and requirements detailed in Tab.~\ref{tab:cutflow}.}
\label{tab:xsecAfterCuts}
\end{center}
\end{table}

Finally after imposing all cuts, we are left with the cross-sections listed in Tab.~\ref{tab:xsecAfterCuts}. For the case of $\lambda_{\textrm{SM}}$, we obtain a signal over background ratio of $S/B \sim 0.14$ at leading order.  For the design luminosity of 30/ab, this translates into a statistical significance, $S/\sqrt{B} \sim 6.3$ upon assuming no systematic uncertainties. Finally we feed these results into a log-likelihood CLs hypothesis test assuming the SM as the null hypothesis and also assuming no systematic uncertainties. At 68\% (95\%) confidence level, we find

\begin{eqnarray}
-3.09 < \kappa_{\lambda} < 2.44 \; (-3.60 < \kappa_{\lambda} < 3.16) \; \; \; \; \textrm{3/ab} \nonumber\\
-2.56 < \kappa_{\lambda} < 1.64 \; (-2.83 < \kappa_{\lambda} < 2.06)  \; \; \; \; \textrm{30/ab}
\end{eqnarray}

Assuming a flat 5\% (10\%) systematic uncertainty, the 68\% confidence level limits change to

\begin{eqnarray}
-3.20 < \kappa_{\lambda} < 2.60 \; (-3.43 < \kappa_{\lambda} < 2.92) \; \; \; \; \textrm{3/ab} \nonumber\\
-2.89 < \kappa_{\lambda} < 2.15 \; (-3.27 < \kappa_{\lambda} < 2.70) \; \; \; \; \textrm{30/ab}.
\end{eqnarray}

Lastly, we also perform the same test on the $t\bar{t}hh$ four point vertex. Upon resorting to a model independent bound on the coupling, we obtain the following bounds at 68\% (95\%) confidence level~\footnote{Upon considering an inclusive analysis, Ref.~\cite{Azatov:2015oxa} obtains $\kappa_{t\bar{t}hh} \in [-1.72, 1.15]$ TeV$^{-1}$ at 68\% C.L. at 3/ab. Whereas, upon considering the $m_{hh}$ variable, they obtain stronger bounds, $|\kappa_{t\bar{t}hh}| < 0.14$ Tev$^{-1}$ at 68\% C.L. for the same integrated luminosity. Further optimisation is thus possible for the $t\bar{t}hh$ channel.}

\begin{eqnarray}
-0.53 \; \textrm{TeV}^{-1} < \kappa_{t\bar{t}hh} < 0.89 \; \textrm{TeV}^{-1} \; (-0.81 \; \textrm{TeV}^{-1} < \kappa_{t\bar{t}hh} < 1.17 \; \textrm{TeV}^{-1})  \; \; \; \; \textrm{3/ab} \nonumber\\
-0.25 \; \textrm{TeV}^{-1} < \kappa_{t\bar{t}hh} < 0.61 \; \textrm{TeV}^{-1} \; (-0.39 \; \textrm{TeV}^{-1} < \kappa_{t\bar{t}hh} < 0.75 \; \textrm{TeV}^{-1})  \; \; \; \; \textrm{30/ab}.
\end{eqnarray}

Upon considering a 5\% (10\%) systematic uncertainty, the 68\% confidence level limits become

\begin{eqnarray}
-0.59 \; \textrm{TeV}^{-1} < \kappa_{t\bar{t}hh} < 0.95 \; \textrm{TeV}^{-1} \; (-0.71 \; \textrm{TeV}^{-1} < \kappa_{t\bar{t}hh} < 1.07 \; \textrm{TeV}^{-1})  \; \; \; \; \textrm{3/ab} \nonumber\\
-0.43 \; \textrm{TeV}^{-1} < \kappa_{t\bar{t}hh} < 0.78 \; \textrm{TeV}^{-1} \; (-0.63 \; \textrm{TeV}^{-1} < \kappa_{t\bar{t}hh} < 0.99 \; \textrm{TeV}^{-1})  \; \; \; \; \textrm{30/ab}.
\end{eqnarray}

In order to see if we can gain additional sensitivity, we finally perform a multivariate analysis (MVA) with boosted decision trees (BDT) in the TMVA framework~\cite{2007physics3039H} with the following variables: reconstructed masses of both the Higgs bosons, $h_1, h_2$, reconstructed mass of the hadronic top-quark, visible part of the reconstructed leptonic top-quark, transverse momenta of these four objects, transverse momenta of the 6 $b$-tagged jets, hardest two light jets and the isolated lepton, the missing transverse energy and $\Delta R(h_1 h_2/h_i t_h/ h_i t_{\textrm{lep}}^{\textrm{vis}})$, where $i=1,2$, and the total number of jets. We train the SM $t\bar{t}hh$ sample with the dominant backgrounds, \textit{viz.}, $t\bar{t}b\bar{b}b\bar{b}, t\bar{t}h b\bar{b}, t\bar{t} h Z, t\bar{t} Z b\bar{b}$ and $t\bar{t}ZZ$. However, owing to a strong drop in efficiency due to the several reconstructions and requirements, we are left with an inadequate number of Monte Carlo (MC) events for a proper training~\cite{KS}~\footnote{For all the channels except for $t\bar{t}h/Z+$ jets, we started with a million MC events. For the latter two we started with 10 (22) million events for $t\bar{t}h(Z)+$ jets. For all the signal samples, we end up with $\sim 600-700$ MC events after all the cuts. For the various backgrounds however, we end up with $\sim 20-200$ MC events. These are not enough to properly train an MVA.}. However, the various variables involved are mostly indiscernible and a BDT is not efficient in improving the sensitivity. For completeness, we find that our $S/B$ improves to $\sim 0.17$ with the statistical significance increasing to $\sim 6.4$. Thus, we did not pursue the MVA analysis further.

The results obtained above are assuming a non-linear realisation of the EFT. Furthermore, we do not marginalise over the other parameters while quoting the constraints. This may change our results to some degree. This first study is important to show the sensitivity of the complicated $t\bar{t}hh$ channel in the $6b, 1\ell+$ jets final state. It is worth mentioning that even though the constraint on the Higgs self-coupling is weaker than what obtains through the $pp \to hh$ production, the constraint on $\kappa_{t\bar{t}hh}$ is of the same order as the one obtained from the $p p \to hh$ channel as shown in Ref.~\cite{Azatov:2015oxa}. This motivates us to combine multiple double Higgs production modes to constrain these couplings even better. Also from Ref.~\cite{Carvalho:2016rys}, where the couplings are varied one at a time, the $pp \to hh$ channel yields a constraint on $\kappa_{t\bar{t}hh}$ which is at least an order of magnitude weaker than the limit derived here. All these results encourage us to study the $t\bar{t}hh$ channel in other final states at the 100 TeV collider and combine the results with the $p p \to hh$ analyses.

Before concluding, we want to comment on the perturbative order of our calculations.  Ref.~\cite{Frederix:2014hta} evaluated the impact of NLO $K$-factors on the total cross sections for the various double Higgs production processes, which are typically of the order of 25\% for our channel.  While for the signal processes, the NLO corrections are known and calculable with standard tools, this is not true for the significantly more involved background processes, \textit{i.e.}, $t\bar{t}b\bar{b}b\bar{b}$ (QCD), $t\bar{t}h(Z)b\bar{b}$, $t\bar{t}h(Z)+$ jets etc.  Because we want to treat all processes on identical footing, we did not include NLO corrections to the signal subset only.  In addition, it is well known that multi-jet merging when applied correctly, is very well capable of recovering the impact of higher-order corrections on shape observables.  This is supported, \textit{e.g.}, by Figs.\ 4 and 5 in Ref.~\cite{Frederix:2014hta}, where the bin-by-bin $K$-factor for the $p_T$ of the hardest (second-hardest) Higgs boson are constant within 10\% or better at around 0.8 (0.75) for $p_T$ values up to 350 GeV (300 GeV). Larger fluctuations in the tails can be traced back to even higher orders, in this case to the emission of more than one additional parton.  Our study does not focus particularly on the tails of the transverse momentum distributions, and because our discriminatory observables (as listed in Tab.~\ref{tab:cutflow}) are mostly invariant masses, we do not expect any significant changes in shapes due to NLO effects. To account for NLO effects on total cross sections, we add an additional 30\% systematic uncertainty on the total rates, and we find at 68\% CL and at 30 ab$^{-1}$

\begin{eqnarray}
-4.23 < \kappa_{\lambda} < 3.98, \; \; -1.18 \; \textrm{TeV}^{-1} < \kappa_{t\bar{t}hh} < 1.54 \; \textrm{TeV}^{-1}.
\end{eqnarray}

\section{Summary and Conclusions}
\label{sec:summary}

One of the most important tasks after the discovery of the Higgs boson and after studying its couplings with gauge bosons and third-generation fermions is to understand the interactions of the scalar sector underlying electroweak symmetry breaking in more detail. One of its cornerstones is the trilinear interaction of the Higgs boson $\kappa_{\lambda}$. Within most realistic extensions of the Standard Model one does not expect a modified Higgs self-interaction in isolation, but modifications of various couplings, \textit{i.e.} the presence of many additional operators. Such new operators would simultaneously contribute to di-Higgs production processes, \textit{e.g.} $pp \to hh$, and would therefore result in blind directions in a global fit. To over-constrain the system of operators expected in extensions of the Standard Model it is consequently of crucial importance to measure as many multi-Higgs final states as possible.

We have revisited the sensitivity of the process $pp \to t\bar{t}hh$ at a future circular collider with $\sqrt{s}=100$ TeV. To take into account deformations of the Standard Model, we varied the two operators $\kappa_{\lambda}\lambda_{\textrm{SM}} h^3 $ and $\kappa_{t\bar{t}hh} (\bar{t}_L t_R h^2 + \hc)$ independently. While the signal cross section increases by a factor of 75 between $\sqrt{s}=14$ TeV and $\sqrt{s}=100$ TeV, the total background before cuts increases by $\sim 40$. Each background has a different enhancement factor and this total factor becomes $\sim 80$ if we don't take into account the $W^{\pm} +$jets backgrounds which are completely negligible after the analysis. Hence, we surveyed a comprehensive list of backgrounds and found that the two operators can be constrained to $-2.14 < \kappa_{\lambda} < 1.60$ and $-0.25 \; \textrm{TeV}^{-1} < \kappa_{t\bar{t}hh} < 0.61 \; \textrm{TeV}^{-1}$ for an integrated luminosity of $30/\mathrm{ab}$ at 68\% C.L. assuming zero systematics. 

While the limit on $k_\lambda$ is not competitive to the predicted limits from the processes $pp \to hh$ \cite{Contino:2016spe, Azatov:2015oxa,Papaefstathiou:2015iba,Goncalves:2018yva} or $pp \to hhj$ \cite{Banerjee:2018yxy} for a 100 TeV collider, the fact that both coefficients $\kappa_{\lambda}$ and $\kappa_{t\bar{t}hh}$ are contributing at tree-level to $t\bar{t}hh$ production means that this process is of significant importance to include in an agnostic global fit for non-linear EFT parameters along with the parameters affecting the $t\bar{t}h$, $ggh$ and $gghh$ vertices. However, it is important to realise that these vertices can already be constrained from the gluon-fusion and VBF productions ($pp \to hh, hhj, hhjj$) and the $t\bar{t}hh$ process will help in disentangling these further owing to having different linear combinations in the couplings.

\subsubsection*{Acknowledgements}
We thank Silvan Kuttimalai for help with Sherpa. We also thank Shilpi Jain, Jonas Lindert and Marek Schoenherr for helpful discussions during various stages of this work. SB is supported by a Durham Junior Research Fellowship CO-FUNDed by Durham University and the European Union, under grant agreement number 609412.

\bibliography{refs}    

\end{document}